\documentclass[aps,prl,twocolumn,groupedaddress,showpacs]{revtex4}
\usepackage{amssymb}
\usepackage{graphicx}
\usepackage{hyperref}

\begin{document}

\title{Heralded Entanglement between Atomic Ensembles:\\
Preparation, Decoherence, and Scaling}
\author{J. Laurat, K. S. Choi, H. Deng, C. W. Chou, and H. J. Kimble}
\affiliation{Norman Bridge Laboratory of Physics 12-33, California Institute of
Technology, Pasadena, California 91125, USA}
\date{\today}

\begin{abstract}
Heralded entanglement between collective excitations in two atomic
ensembles is probabilistically generated, stored, and converted to
single photon fields. By way of the concurrence, quantitative
characterizations are reported for the scaling behavior of
entanglement with excitation probability and for the temporal
dynamics of various correlations resulting in the decay of
entanglement. A lower bound of the concurrence for the collective
atomic state of $0.9\pm 0.3$ is inferred. The decay of entanglement
as a function of storage time is also observed, and related to the
local dynamics.
\end{abstract}

\maketitle

Beyond a fundamental significance, quantum control of entanglement between
material systems is an essential capability for quantum networks and
scalable quantum communication architectures \cite{zoller05,Briegel98}. In
recent years, significant advances have been achieved in the control of the
quantum states of atomic systems, including entanglement of trapped ions
\cite{Haffner05,Langer05} and between macroscopic spins \cite%
{Julsgaard01}. By following the seminal paper of Duan, Lukin, Cirac
and Zoller (\textit{DLCZ}) \cite{duan01}, entanglement between
single collective excitations stored in two remote atomic ensembles
has also been demonstrated \cite{chou05}. In the \textit{DLCZ}
protocol, entanglement is created in a probabilistic but heralded
way from quantum interference in the measurement process
\cite{dicke,Cabrillo99,knight}. The detection of a photon from one
or the other atomic ensemble in an indistinguishable fashion results
in an entangled state with one collective spin excitation shared
coherently between the ensembles. Such entanglement has been
critical for the initial implementation of functional quantum nodes
for entanglement distribution \cite{chou07}, for the investigation
of entanglement swapping \cite{laurat07} and for light-matter
teleportation \cite{pan3}.

Because of the relevance to quantum networking tasks, it is important to
obtain detailed characterizations of the physical processes related to the
creation, storage, and utilization of heralded entanglement. Towards this
end, significant advances have been demonstrated in the generation of
photon-pairs \cite{chou04,Balic05} and the efficient retrieval of collective
excitation \cite{laurat06,thompson06}. Moreover, decoherence processes for a
single atomic ensemble in the regime of collective excitation have been
investigated theoretically \cite{felinto05} and a direct measurement of
decoherence for one stored component of a Bell state recently performed \cite%
{deRiedmatten06}. However, to date no direct study has been reported
for the decoherence of an entangled system involving two distinct
atomic ensembles, which is a critical aspect for the implementation
of elaborate protocols \cite{lukin,pan,pan2}. The decoherence of
entanglement between ensembles has been shown in recent setups,
through the decay of the violation of a Bell inequality
\cite{chou07} and the decay of the fidelity of a teleported state
\cite{pan3}. However, a quantitative analysis was not provided since
these setups involved many others parameters, such as phase
stability over long distances.

In this Letter, we report measurements that provide a detailed and
quantitative characterization of entanglement between collective
atomic excitations. Specifically, we determine the concurrence $C$
\cite{wooters98} as a function of the normalized degree of
correlation $g_{12}$ \cite{laurat06}
for the ensembles, including the threshold $g_{12}^{(0)}$\ for entanglement $%
(C>0)$. We also map the decay of the concurrence $C(\tau )$ as a
function of storage time for the entangled state, and interpret this
decay by measuring the local decoherence on both ensembles taken
independently. Compared to Ref. \cite{chou05}, these observations
are made possible by a new system that requires no active phase
stability and that implements conditional control for the
generation, storage, and readout of entangled atomic states.

\begin{figure}[t!]
\includegraphics[width=.85\columnwidth]{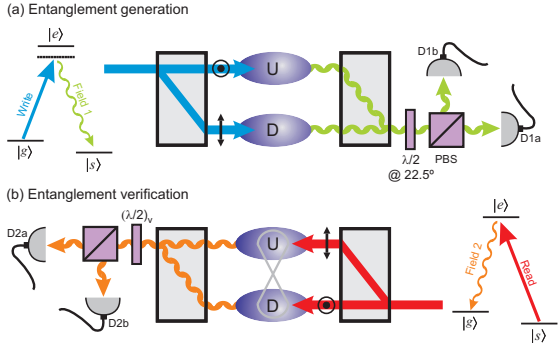}
\caption{(a) A weak write pulse is split into two paths separated by
$1$ mm
and excites simultaneously two atomic samples, $%
U,D $. The resulting fields $1_{U,D}$ are combined at the polarizing
beamsplitter (PBS) and sent to the single-photon detectors $D_{1a,1b}$. A
detection event at $D_{1a}$ or $D_{1b}$ heralds the creation of entanglement. (b) After a storage time $\protect%
\tau $, entanglement is verified by mapping the atomic state to
propagating fields $2_{U,D}$ by way of read pulses. Tomography is
then achieved in two steps, as described in the text.
The atomic cloud is initially prepared in the ground state $|g\rangle $. $%
\{|g\rangle ,|s\rangle ,|e\rangle \}$ denote the levels $%
\{|6S_{1/2},F=4\rangle ,|6S_{1/2},F=3\rangle ,|6P_{3/2},F=4\rangle
\}$. Note that the fields $1_{U,D}$ and $2_{U,D}$ are detected with
a small angle relative to the classical beams, which is not
represented here for the sake of simplicity.} \label{setup}
\end{figure}

Our experiment is illustrated in Fig. \ref{setup}. A single cloud of cesium
atoms in a magneto-optical trap is used; two ensembles are defined by
different optical paths $1$ mm apart \cite{Matsu04,chou07}. This separation
is obtained by the use of birefringent crystals close to the cloud, which
separate orthogonal polarizations \cite{chouthesis}. At $40$ Hz, the trap
magnetic field is switched off for $7$ ms. After waiting $3$ ms for the
magnetic field to decay, the two samples are simultaneously illuminated with
$30$ ns-long and $10$ MHz red-detuned write pulses, at a rate of $1.7$ MHz.
Given the duty cycle of the experiment, the effective rate is $180$ kHz.
Spontaneous Raman scattered fields induced by the write beams are collected
into single-mode fibers, defining for each ensemble optical modes that we
designate as fields $1_{U,D}$ with 50 $\mu m$ waist and a 3$^{\circ } $
angle relative to the direction of the write beams \cite{Balic05,laurat06}.
Fields $1_{U,D}$ are frequency filtered to block spontaneous emission from
atomic transitions $|e\rangle \rightarrow |g\rangle $, which do not herald
the creation of a collective excitation. After this stage, and before
detection, fields $1_{U,D}$ are brought to interfere on a polarizing
beam-splitter. A detection event at $D_{1a,1b}$ that arises
indistinguishably from either of the fields $1_{U,D}$ projects the atomic
ensembles into an entangled state where, in the ideal case, one collective
excitation is coherently shared between the $U,D$ ensembles \cite%
{duan01,chou05}.

In the ideal case of small excitation probability, the atom-field 1 joint
state can be written for each ensemble:
\begin{equation}
|\Psi \rangle =|0_{a}\rangle |0_{1}\rangle +\sqrt{\chi }|1_{a}\rangle
|1_{1}\rangle +O(\chi )\text{\textit{\ } ,}  \label{field1atomes}
\end{equation}%
with $|n_{1}\rangle $ the state of the field 1 with $n$ photons and $%
|n_{a}\rangle $ the state of the ensemble with $n$ collective
excitations. Upon a detection event at $D_{1a,1b}$, in the ideal
case, the atomic state is projected into
\begin{equation}
|\Psi _{U,D}\rangle =\frac{1}{\sqrt{2}}(|0_{a}\rangle _{U}|1_{a}\rangle
_{D}\pm e^{i\theta }|1_{a}\rangle _{U}|0_{a}\rangle _{D})+O(\sqrt{\chi}),
\label{state}
\end{equation}%
where $|0_{a}\rangle _{U,D},|1_{a}\rangle _{U,D}$ refers to the two
ensembles $U,D$ with $0,1$ collective excitations, respectively \cite{duan01}%
. The $\pm $ sign is set by the detector that records the heralding event.
The overall phase $\theta $ is the sum of the phase difference of the write
beams at the $U$ and $D$ ensembles and the phase difference acquired by
fields $1$ in propagation from the ensembles to the beamsplitter. To achieve
entanglement, this phase must be constant from trial to trial \cite{vanEnk06}%
. In order to meet this requirement, the initial demonstration
reported in \cite{chou05} employed auxiliary fields to achieve
active stabilization for various phases for two ensembles located in
distinct vacuum apparatuses. By contrast, in our current setup (Fig.
\ref{setup}a), $\theta $ is determined only by the differential
phase for the two paths with orthogonal polarizations defined by the
birefringent crystals \cite{chouthesis}; our small setup has
sufficient passive stability without need of adjustment or
compensation as the phase does not change by more than a few degrees
over 24 hours.

To verify operationally entanglement between the $U,D$ ensembles,
the respective atomic states are mapped into photonic states by
applying simultaneously read pulses in the configuration introduced
in Ref. \cite{Balic05}, as depicted in Figure \ref{setup}b. The
delocalized atomic excitation is retrieved with high efficiency
thanks to  collective enhancement \cite{duan01,laurat06} and, in the
ideal case, $|\Psi
_{U,D}\rangle $ would be mapped directly to the photonic state of fields $%
2_{U,D}$ with unity efficiency and no additional components.
Stability for the phase difference of the read beams and of fields
$2_{U,D}$ is also required in this process; it is again achieved by
the passive stability of our current scheme \cite{chouthesis}. Since
entanglement cannot be increased by local operations
\cite{Bennett96}, the entanglement for the atomic state will always
be greater than or equal to that measured for the light fields.

A model-independent determination of entanglement based upon quantum
tomography of the fields $2_{U,D}$ has been developed in Ref.
\cite{chou05}. The model consists of reconstructing a density
matrix, $\tilde{\rho}_{2_U,2_D}$, obtained from the full density
matrix by restriction to the subspace with no more than one photon
per mode. We also assume that all off-diagonal elements between
states with different numbers of photons vanish. The model thus
leads to a lower bound for entanglement. As detailed in Ref. \cite{chou05}, $%
\tilde{\rho}_{2_U,2_D}$ can be written in the photon-number basis $|n\rangle
|m\rangle $ with $\{n,m\}=\{0,1\}$ as follows:%
\begin{equation}
\tilde{\rho}_{2_U,2_D}=\frac{1}{P}\left(
\begin{array}{cccc}
p_{00} & 0 & 0 & 0 \\
0 & p_{01} & d & 0 \\
0 & d^{\ast } & p_{10} & 0 \\
0 & 0 & 0 & p_{11}%
\end{array}%
\right) \text{ .}  \label{rho}
\end{equation}%
Here, $p_{ij}$ is the probability to find $i$ photons in mode $2_{U}$ and $j$
in mode $2_{D}$; $d$ is the coherence term between the $|1\rangle |0\rangle $
and $|0\rangle |1\rangle $ states; and $P=p_{00}+p_{01}+p_{10}+p_{11}$. From
$\tilde{\rho}_{2_U,2_D}$, one can calculate the concurrence $C$, which is a
convenient monotone measurement of entanglement ranging from $0$ for a
separable state to $1$ for a maximally entangled state \cite{wooters98}:
\begin{equation}
C=max(0,C_0)\,\,
\text{with}\,\,C_0=\frac{1}{P}(2|d|-2\sqrt{p_{00}p_{11}}). \label{C}
\end{equation}%
In the regime of low excitation and detection probabilities in which
the experiment is performed, the vacuum $p_{00}$ can be approximated
by $p_{00}\sim 1-p_{c}$, while the terms $p_{01}$ and $p_{10}$ are
given by $p_{10}=p_{01}=p_{c}/2$. $p_{c}$ is the conditional
probability of detecting a photon in field $2$ from one ensemble
following a detection event for field $1$.

Experimentally, we reconstruct $\tilde{\rho}_{2_U,2_D}$ and then calculate $C
$ by using two configurations for the detection of fields $2_{U,D}$,
corresponding to two settings of the $(\lambda /2)_{v}$ waveplate shown in
Fig. \ref{setup}b. The diagonal elements of $\tilde{\rho}_{2_U,2_D}$ are
determined from measurements of the photon statistics for the separated
fields $2_{U},2_{D}$, i.e., by detecting independently each field. To access
the coherence term $d$, fields $2_{U,D}$ are coherently superimposed and the
count rates from the resulting interference are recorded as a function of
the relative phase between the $2_{U,D}$ fields. It can be shown that $%
d\simeq V(p_{10}+p_{01})/2\sim Vp_{c}/2$ \cite{chou05}, where $V$ is
the visibility of the interference fringe.

To investigate the scaling of entanglement with excitation
probability $\chi$, we determine $C$ for various values of $\chi$
for fixed memory time $\tau
=200$ ns. As $\chi$ increases, higher order terms in the expansion of Eq. (%
\ref{state}) cannot be neglected, precisely as in parametric down
conversion. A convenient parameter to assess the excitation regime
of each ensemble is the normalized intensity cross correlation
function $g_{12}$ between field $1$ and field $2$ \cite{laurat06},
defined as $g_{12}=p_{12}/(p_1p_2)$ with $p_{12}$ the joint
probability for detection events from field 1 and 2 in a given trial
and $p_i$ the probability for unconditional detections in field $i$.
In the ideal case, this function is related to the excitation
probability $\chi$ by $g_{12}=1+1/\chi
$, where $g_{12}>2$ defines the nonclassical border in the ideal case \cite%
{chou04} and $g_{12}\gg 2$ being the single-excitation regime for the
ensembles.
\begin{figure}[t]
\includegraphics[width=.8\columnwidth]{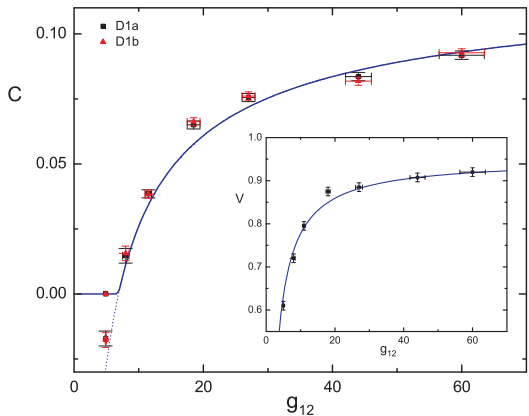}
\caption{Concurrence $C$ as a function of the normalized cross
correlation
function $g_{12}$, for the two possible heralding events (detection at $%
D_{1a}$ or $D_{1b}$). The solid line is obtained from Eq. (\protect\ref{Conc}%
) with the fitted overlap (see inset) and assuming an
independently-measured retrieval efficiency at 13.5$\%$. The dotted
line corresponds to $C_0$. Inset: Average visibility of the
interference fringe between the two field-2 modes. The solid line is
a fit using the expression given by Eq. (\protect\ref{Visibility}),
with the overlap $\protect\xi $ fitted to $0.95\pm 0.01$.}
\label{g12}
\end{figure}

Expressing the two-photon component for the two ensembles as $p_{11}= \chi
p_{c}^{2}\sim p_{c}^{2}/g_{12}$, we rewrite the concurrence as:
\begin{equation}
C\simeq max[0,p_{c}(V-2\sqrt{(1-p_{c})/g_{12}})]\text{ ,}  \label{Conc}
\end{equation}%
where $g_{12}$ is for either ensemble alone, with $g_{12}^{(U)}=g_{12}^{(D)}%
\equiv g_{12}$ assumed. The visibility $V$ can be expressed in terms of $%
g_{12}$ as the higher order terms act as a background noise. With
$(1/2)p_1p_2$ a good estimation for the background, the visibility
can be written as \cite{chouthesis}
\begin{equation}
V\simeq \xi \,\frac{p_{12}-p_1p_2}{p_{12}+p_1p_2}= \xi\,
\frac{g_{12}-1}{g_{12}+1}\text{ ,}  \label{Visibility}
\end{equation}%
where $\xi $ is the overlap between fields $2_{U,D}$ \cite{felinto06}. In
the limit of near zero excitation, as $g_{12}$ goes to infinity, the
concurrence reaches its asymptotic value given by the retrieval efficiency $%
\xi p_{c}$ \cite{h}.

Fig. \ref{g12} presents our measurements of the concurrence $C$ as a
function of $g_{12}$. As the excitation probability is decreased,
$g_{12}$ increases as does the entanglement. The threshold to
achieve $C>0$ is found to be $g_{12}^{(0)}\simeq 7$, corresponding to a probability $%
p\simeq 1.2\times 10^{-2}$ per trial for the creation of the heralded
entangled state and to a preparation rate $\sim 2$ kHz. Note that $C=0$ (or $%
C$ not greater than zero) does not imply that there is no
entanglement, only that any possible entanglement is not detected by
our protocol, which provides a lower bound for the entanglement.
More importantly, in an infinite dimensional Hilbert space,
entangled states are dense in the set of all states \cite{Eisert02},
so that zero entanglement is not provable for an actual experiment
by way of the concurrence.

To confirm the model leading to Eq. (\ref{Visibility}), the inset
gives the measured visibility $V$ as a function of $g_{12}$. The
solid line is a fit according to Eq. (\ref{Visibility}) with free
parameter $\xi $,
leading to an overlap $\xi =0.95\pm 0.01$, in agreement with the value $%
\xi =0.98\pm 0.03$ obtained from an independent two-photon
interference measurement. With the fitted value of $\xi $ and with
the
independently determined value of the conditional probability $%
p_{c}=0.135\pm 0.005$ from measurements performed on each ensemble
separately, we compare our measurements of $C$ with the prediction of Eq. %
\ref{Conc} (solid line in Fig. \ref{g12}) and find good agreement.
\begin{table}[b]
\caption{Diagonal elements and concurrence of the density matrices
for fields $2_{U,D}$, without and with correction for propagation
losses and detection efficiencies. The last column provides the
estimated elements and
concurrence for the atomic state by considering the readout efficiency $%
\protect\eta $. $g_{12}=60\pm 4$.}{\footnotesize
\begin{ruledtabular}
\begin{tabular}{lccc}
&$\tilde{\rho}_{2_{U},2_{D}}$&$\tilde{\rho} _{2_{U},2_{D}}^{output}$&$\tilde{\rho} _{U,D}$\\
$p_{00}$&$0.864\pm0.001$&$0.54\pm0.08$&$0\pm0.3$\\
$p_{10}$&$( 6.47\pm0.02)\times 10^{-2}$&$(22\pm4)\times 10^{-2}$&$0.5\pm0.15$\\
$p_{01}$&$(7.07\pm0.02)\times 10^{-2}$&$(24\pm4)\times 10^{-2}$&$0.5\pm0.15$\\
$p_{11}$&$(2.8\pm0.2)\times 10^{-4}$&$(3\pm2)\times 10^{-3}$&$0.015\pm0.025$\\
$C$&$0.092\pm0.002$&$0.35\pm 0.1$&$0.9\pm0.3$\\
\end{tabular}
\end{ruledtabular}}
\label{table}
\end{table}

Table I provides the diagonal elements of the density matrix $\tilde{\rho}%
_{2_{U},2_{D}}$ and the concurrence for the case $g_{12}=60\pm 4$
corresponding to a probability to create entanglement $p=9\times
10^{-4}$ per trial (160 Hz). A value $C=0.092\pm 0.002$ is directly
measured at detectors $D_{2a},D_{2b}$ without correction. By way of
the independently determined propagation and detection efficiencies,
we infer the density matrix $\tilde{\rho}_{2_{U},2_{D}}^{output}$ for fields $%
2_{U},2_{D}$ at the output of the ensembles, from which we obtain a
concurrence $C_{2_{U},2_{D}}^{output}=0.35\pm 0.1$. This value
exceeds the published state of the art by two orders of magnitude
\cite{chou05}. This leap underlines the progress obtained in terms
of suppression of the two-photon component and achievable retrieval
efficiency over the past year \cite{laurat06,thompson06}. Finally,
by way of the conditional readout efficiency $\eta =45\pm 10\%$ for
mapping of quantum states of the $U,D$ ensembles to the fields
$2_{U},2_{D}$, we estimate the density matrix $\tilde{\rho}_{U,D}$
and the concurrence $C_{U,D}=0.9\pm 0.3$ for the collective atomic
state. We emphasize that $C_{U,D}$ is an estimate determined from
the model developed in Ref. \cite{chou04} where the fields at the
output of the MOT consist of a two-mode squeezed state plus
background fields in coherent states.

\begin{figure}[t!]
\includegraphics[width=.83\columnwidth]{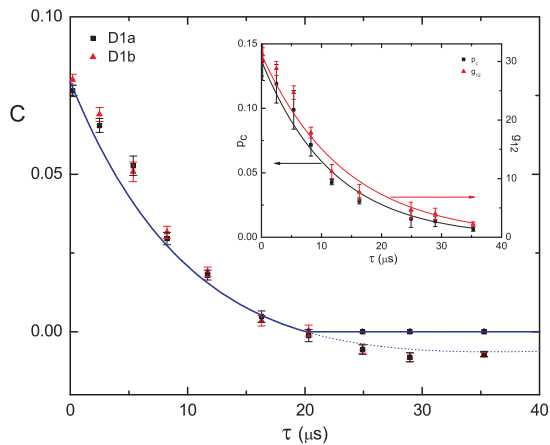}
\caption{Concurrence $C$ as a function of the storage time $%
\protect\tau $. The solid line is obtained from Eq.
(\protect\ref{Conc}) assuming the fitted exponential decays, given
in the inset, of the individual parameters $p_{c}$ and $g_{12}$
measured independently. The dotted line corresponds to $C_0$.}
\label{storage}
\end{figure}
Turning then to a characterization of the decay of entanglement with
storage time $\tau $, we present in Fig. \ref{storage} measurements
of concurrence $C(\tau )$ for fixed excitation probability
$p=1.6\times 10^{-3}$ corresponding to $g_{12}=30$ at $\tau =200$
ns. $C>0$ for $\tau \lesssim 20$ $\mu s$, providing a lower bound
for the lifetime of entanglement of the ensembles corresponding to
$4$km propagation delay in a fiber.

To investigate the dynamics in Fig. \ref{storage}, the inset shows
the decay of the average $g_{12}$ and conditional probability
$p_{c}$ for the ensembles taken independently. Such local
decoherence has been investigated as the result of inhomogeneous
broadening of the Zeeman ground states due to residual magnetic
fields \cite{felinto05,deRiedmatten06,kuzmich05}. Our current
measurement shows the effect of this local decoherence on the
entanglement of the joint system of the ensembles. For this purpose,
our measurements of $C$ are superposed with a line $C(\tau )$ given
by Eq. (\ref{Conc}), where the fitted exponential decay for
$p_{c}(\tau ),g_{12}(\tau )$ (with similar decay $\simeq 13$ $\mu
s$) and the overlap $\xi $ determined in Fig. \ref{g12} are
employed. The agreement evidenced in Fig. \ref{storage} confirms the
principal role of local dephasing in the entanglement decay.

In conclusion, we have reported a detailed study of the behavior of
entanglement between collective excitations stored in atomic
ensembles, including the dependence of the concurrence on the degree
of excitation and the quantitative relationship of local decoherence
to entanglement decay. The temporal dynamics reveal a finite-time
decay, with separability onset for storage time $\tau \sim 20$ $\mu
s$. From a more general perspective, the inferred concurrence for
the collective atomic state, $C_{U,D}=0.9\pm 0.3$, is comparable to
values obtained for entanglement in the continuous variable regime
\cite{laurat05} and for entanglement of the discrete internal states of trapped ions \cite%
{Haffner05,Langer05}.

We acknowledge our collaboration with S. J. van Enk and discussions
with D. Felinto and H. de Riedmatten. This research is supported by
the DTO and by the NSF. J.L. acknowledges financial support from the
EU (Marie Curie fellowship) and H.D. from the CPI at Caltech.


\begin{thebibliography}{99}
\bibitem{h} An alternative approach is to determine the suppression of
two-photon events relative to the square of the probability of
single photon events for the fields 2, $h\equiv
p_{11}/(p_{10}p_{01})$. $h<1$ is a necessary condition for
entanglement \cite{chou05,chou07,laurat07}. Here, for $g_{12}=60$,
$h=0.06$.
\end{thebibliography}
\end{document}